\documentclass[a4paper,11pt]{article}
\usepackage{pos}
\usepackage{xcolor}

\def\to{\rightarrow}

\def\be{\begin{equation}}
\def\ee{\end{equation}}      
\def\bal{\begin{align}}
\def\ba{\begin{array}}
\def\ea{\end{array}}
\def\bmat{\begin{pmatrix*}}
\def\emat{\end{pmatrix*}}
\def\bdet{\begin{vmatrix}}
\def\edet{\end{vmatrix}}

\def\ph{\phantom}

\def\ie{{\em i.e.}, }

\def\red{\color{red}}

\title{Implications of Recent Experimental \& Theoretical Results on Electroweak Precision Tests}
\ShortTitle{Electroweak Precision Tests}

\author*{Jens Erler}

\affiliation{PRISMA$^+$ Cluster of Excellence, Institute for Nuclear Physics, \\
Johannes Gutenberg-University, 55099 Mainz, Germany}

\affiliation{Helmholtz-Institut Mainz, \\
Johannes Gutenberg-Universität, 55099 Mainz, Germany}

\emailAdd{erler@uni-mainz.de}

\abstract{I review the results of a recent global fit to electroweak precision data. 
Particular attention is devoted to the landscape of determinations of the weak mixing angle,
recent results on basic properties of the electroweak gauge bosons,
and the implications of vacuum polarization on the scale dependences of the electromagnetic coupling and the weak mixing angle,
as well as the anomalous magnetic moment of the muon.}

\FullConference{Proceedings of the Corfu Summer Institute 2024 "School and Workshops on Elementary Particle Physics and Gravity" (CORFU2024)\
12 - 26 May, and 25 August - 27 September, 2024\\
Corfu, Greece\\}

 \tableofcontents

\begin{document}
\maketitle
\section{The over-constrained SM}
After half a century of tests and scrutiny, the Standard Model~\cite{Glashow:1961tr,Weinberg:1967tq} (SM) of elementary particle physics still stands its ground.
It is now clear that the SM is correct to leading order (tree level) and is also fully consistent with all relevant experimental observations
when first order corrections are taken into account.
This means that physics beyond the SM can only reveal itself by a small perturbation (or else at very high energies),
and in turn that the SM needs to be over-constrained so that small inconsistencies may eventually turn up.
Presently, there remain some tensions in the anomalous magnetic moment of the muon, the $W$ boson mass, $M_W$,
and the unitarity of the first row of the CKM matrix~\cite{Gorchtein:2023naa}, but the SM continues to be in excellent shape.

As one of the consequences, the SM predictions to high-precision observables need to be understood with uncertainties at a level that should 
be small or ideally even negligible compared to the experimental errors. 
The higher the precision becomes, the more physics issues enter which need to be addressed before a measurement can be properly interpreted.
While this can be an obstacle when only one observable is considered at a time, it may become a feature in global analyses,
where the observables are being viewed from the perspectives of different subfields of particle, nuclear and atomic physics at the same time.
In this sense, one physicists uncertainty may be an opportunity for another. 

Also, the free SM parameters need to be determined with a corresponding accuracy. 
In practice this is done in global electroweak (EW) fits within the SM or allowing parameters describing new physics beyond it. 
The relevant observables can be classified into four categories:
\begin{description}
\item[$Z$-pole observables] These are the $Z$ lineshape observables~\cite{ALEPH:2005ab} (the mass and the width of the $Z$, $M_Z$ and $\Gamma_Z$, 
and the hadronic peak cross section), as well as cross-section ratios and various cross-section asymmetries.
\item[High-energy precision observables] The mass and the width of the $W$, $M_W$ and $\Gamma_W$, 
and the top quark mass $m_t$ are derived from the hadron colliders Tevatron and LHC, while the Higgs mass $M_H$ 
has been measured only at the LHC~\cite{ATLAS:2023oaq,CMS:2023ogh} (there are currently only weak and indirect constrains on its width $\Gamma_H$).
\item[Intermediate-energy precision observables] Other necessary quantities~\cite{ParticleDataGroup:2024cfk} are the quark masses $m_b$ and $m_c$,
the strong coupling $\alpha_s$, as well as the vacuum polarization entering several key precision observables (see Section~\ref{HVP}).
\item[Low-energy precision observables] Observables characterized by typical momentum transfers below the hadronic scale~\cite{Kumar:2013yoa},
such as from parity-violating electron scattering (PVES) or atomic parity violation (APV), are complementary to those from the energy frontier.
\end{description}
Some (combinations) of these observables determine the SM parameters.
Additional ones then over-constrain the SM and may be used to look for new physics.

In this context, two observables, $M_W$ and the weak mixing angle $\sin^2\theta_W$~\cite{Glashow:1961tr} are of particular importance,
since they are ultra-precise and at the same time, they can be calculated within the SM.
To set the stage, I start with three equations~\cite{Weinberg:1967tq},
\begin{equation}
\sin^2\theta_W = \frac{g'^2}{g^2 + g'^2} = 1 - \frac{M_W^2}{M_Z^2} = \frac{\pi\alpha}{\sqrt{2} G_F M_W^2}
\end{equation}
where $g$ and $g'$ are the $SU(2)_L$~\cite{Yang:1954ek} and $U(1)_Y$ gauge couplings, 
and $\alpha$ and $G_F$ denote the fine structure constant and the Fermi constant~\cite{Fermi:1934hr}.
These are tree-level relations which are modified by various radiative corrections~\cite{Hollik:1988ii}, 
\begin{equation}
\frac{\sin^2\theta_W^\ell}{\red 1 + \Delta\hat k} = \frac{g'^2}{g^2 + g'^2} = 1 - \frac{{\red (1 - \Delta\hat\rho)}M_W^2}{M_Z^2} = 
\frac{\pi\alpha}{{\red (1 - \Delta\hat r)}\sqrt{2} G_F M_W^2}
\label{EWheart}
\end{equation}
where $\sin^2\theta_W^\ell$~\cite{Gambino:1993dd} is an effective parameter entering the vector coupling $v_\ell^Z$ of the $Z$ boson to leptons.
The large value of $m_t$ drives the parameter $\Delta\hat\rho \propto m_t^2/M_W^2$~\cite{Veltman:1977kh} to values of order $1\%$
compared to the few per mille level of typical EW corrections.
The parameter $\Delta\hat r$~\cite{Sirlin:1989uf} is numerically the largest (close to 7\%), 
as it contains the renormalization group evolution (running) of the electromagnetic coupling from the Thomson limit to the $Z$ boson mass scale.
Eq.~\eqref{EWheart} is at the heart of any EW fit, and --- clearly --- 
one needs to understand radiative correction parameters very precisely to be able to isolate new physics.

\begin{figure}[t]
\begin{minipage}{\linewidth}
\centerline{\includegraphics[trim={0 0 0 180},clip,width=1.11\linewidth]{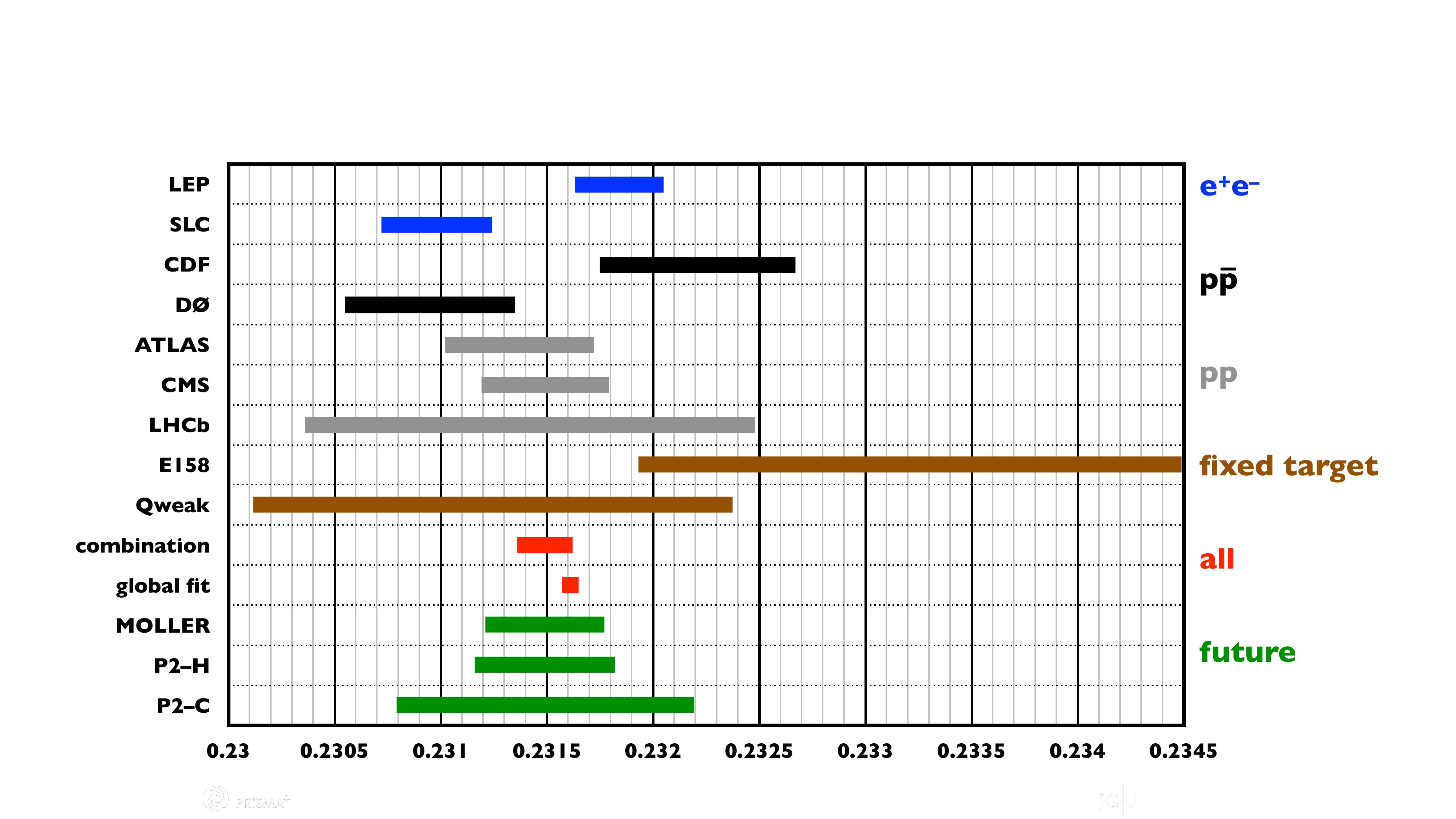}}
\end{minipage}
\caption[]{Determinations of the weak mixing angle $\sin^2\theta_W^\ell$ (assuming leptin universality) 
from the $Z$~boson factories LEP and SLC~\cite{ALEPH:2005ab} (in blue),
from CDF and D\O\ at the $p\bar p$ collider Tevatron~\cite{CDF:2018cnj} (in black), from ATLAS~\cite{ATLAS:2015ihy,ATLAS:2018gqq},
CMS~\cite{CMS:2018ktx,CMS:2024ony} and LHCb~\cite{LHCb:2015jyu} at the proton collider LHC (in gray), 
as well as from fixed target experiments in parity-violating electron scattering
from E158~\cite{SLACE158:2005uay} at SLAC and Qweak~\cite{Qweak:2018tjf} at Jefferson Lab (in brown).
The combination and the result of the global EW fit are also shown (in red).
The last three entries are the projected uncertainties of the PVES experiments MOLLER~\cite{Demiroglu:2024wys} at Jefferson Lab and 
P2~\cite{Becker:2018ggl} at the Mainz Energy-Recovering Superconducting Accelerator (MESA), both currently under construction (in green).}
\label{sin2tw}
\end{figure}

Given the importance of the weak angle for EW physics, we summarize the experimental situation in Figure~\ref{sin2tw}.
The reason for the exquisite precision demonstrated in the figure is that $v_\ell^Z$ is proportional to $1 - 4 \sin^2\theta_W^\ell$ at tree level,
combined with the fact that the numerical value of $\sin^2\theta_W^\ell$ is close to 1/4.
This suppression of the SM prediction also results in a relative enhancement and thus in extra sensitivity to physics beyond the SM (BSM).
The tension between the LEP and SLC results notwithstanding, one generally observes good experimental agreement.
One of the largest sources of uncertainty entering the determinations at the hadron colliders are from parton distribution functions (PDF).
However, the Tevatron, the general purpose detectors at the LHC (ATLAS and CMS), and LHCb are largely sensitive to different aspects of the PDFs,
so as to reduce their mutual correlations.

The SLAC--E158~\cite{SLACE158:2005uay} and Qweak~\cite{Qweak:2018tjf} Collaborations measured the left-right cross section asymmetry,
\be
A_{LR} = \frac{\sigma_L - \sigma_R}{\sigma_L + \sigma_R} \sim \frac{G_F Q^2}{4\pi\alpha} \sim 10^{-4}\, Q^2 \mbox{ [GeV}^2]
\label{ALR}
\ee
in experiments using high-intensity longitudinally polarized electron beams scattering off unpolarized liquid hydrogen targets.
Pure QED effects cancel in the numerator of $A_{LR}$, such that the leading effect is the $\gamma Z$ interference contribution,
selecting the $Z$ vector coupling at lowest EW order.
E158 measured the asymmetry in M\o ller scattering by focussing on scattering angles of only a few mrad in the laboratory frame,
while Qweak measured the $e^- p$ asymmetry at about $8^\circ$.
Since the CEBAF beam energy of 1.149~GeV delivered to Qweak was about 40 times lower than that of the SLC beam used by E158,
the fixed-target kinematics coincidentally resulted in both cases in low-momentum transfers of $Q^2 \approx 0.025$~GeV 
for which $A_{LR} \sim 10^{-7}$ is very small.
The MOLLER Collaboration will improve on E158 by almost a factor of five in precision, 
by exposing its detector to the upgraded 11~GeV CEBAF beam, even though $A_{LR} = 3.3 \times 10^{-8}$ is even smaller.

\begin{figure}[t]
\hspace{18pt}
\begin{minipage}{\linewidth}
\centerline{\includegraphics[trim={0 0 0 30},width=1.08\linewidth]{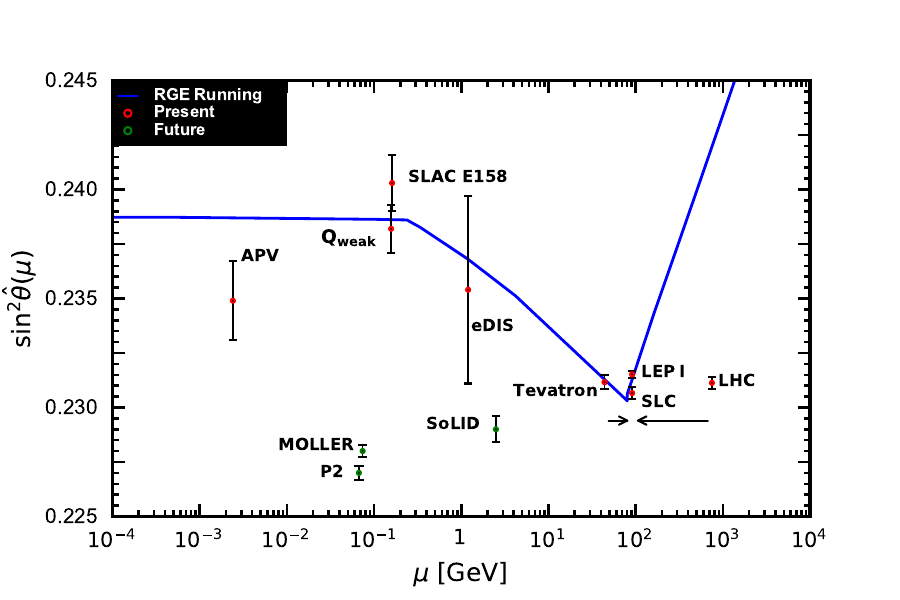}}
\end{minipage}
\caption[]{Determinations of the weak mixing angle $\sin^2\hat\theta_W(\mu)$ in the $\rm \overline{MS}$ scheme as a function of 
the renormalization scale $\mu$, compared to the SM prediction~\cite{Erler:2004in,Erler:2017knj}.
The collider results are all taken close to the $Z$ peak 
but have been displaced for clarity, so they should be shifted horizontally back to $\mu = M_Z$, as indicated.
Very recently, the first purely theoretical evaluation of the running~\cite{Erler:2024lds}, 
\ie without experimental constraints from $e^+ e^-$ annihilation or $\tau$ decays,
has been computed together with the correlation with the running of the electromagnetic coupling to the $Z$ scale, 
$\Delta\alpha(M_Z)$~\cite{Davier:2023cyp}, and with the anomalous magnetic moment of the muon~\cite{Muong-2:2023cdq,Aoyama:2020ynm}. 
The behaviour of an effective definition, $\sin^2\theta_W (Q^2)$, is shown in Ref.~\cite{Czarnecki:2000ic}.}
\label{runningsin2tw}
\end{figure}

In the case of $e^- p$ scattering, the non-trivial proton form factors need to be taken into account,
introducing an additional uncertainty in the extraction of the so-called weak charge of the proton, $Q_W(p) \approx 1 -4 \sin^2\theta_W$. 
Radiative corrections~\cite{Marciano:1982mm,Erler:2003yk}, especially from box $WW$ and $\gamma Z$ box diagrams, 
are also much larger in the proton case.
Indeed, while the vector-coupling of the $Z$ boson is not renormalized in the Thomson limit, 
the $\gamma Z$ box is lifting the $1 -4 \sin^2\theta_W$ suppression~\cite{Gorchtein:2008px} by an amount roughly proportional to the beam energy.
To reduce the uncertainty of the latter, the P2 Collaboration~\cite{Becker:2018ggl} is constructing a 155~MeV superconducting $e^-$
accelerator on the JGU campus in Mainz, Germany, with the precision goal $\Delta\sin^2\theta_W = \pm 0.00035$,
which is close to the precision achieved in the most sensitive observables at the $Z$ factories LEP and SLC.
P2 is also planning to measure $A_{LR}$ on a $^{12}$C target, which is proportional to $\sin^2\theta_W$ and thus does not benefit 
from the suppression by $1 - 4 \sin^2\theta_W$.
On the other hand, cross sections and asymmetry are larger in this case, allowing for a much smaller statistical uncertainty.  
The carbon measurement will likely be limited by polarimetry.

One can also measure $A_{LR}$ in deep-inelastic electron scattering (eDIS) where electrons scatter mainly off individual quarks, 
and where at $Q^2 \sim 1$~GeV the asymmetry is much larger, $A_{LR} \sim 10^{-4}$.
Historically, the first experiment of this kind, E122 at SLAC~\cite{Prescott:1978tm}, established the basic structure of the SM, 
and more recently, the PVDIS Collaboration at Jefferson Lab~\cite{PVDIS:2014cmd} was first to determine 
the axial-vector coupling of the first generation quarks.
The next-generation experiment SoLID~\cite{Tian:2024xum} is expected to achieve a high-precision measurement in the DIS regime.

\begin{figure}[t]
\begin{minipage}{\linewidth}
\centerline{\includegraphics[trim={0 0 0 180},clip,width=1.11\linewidth]{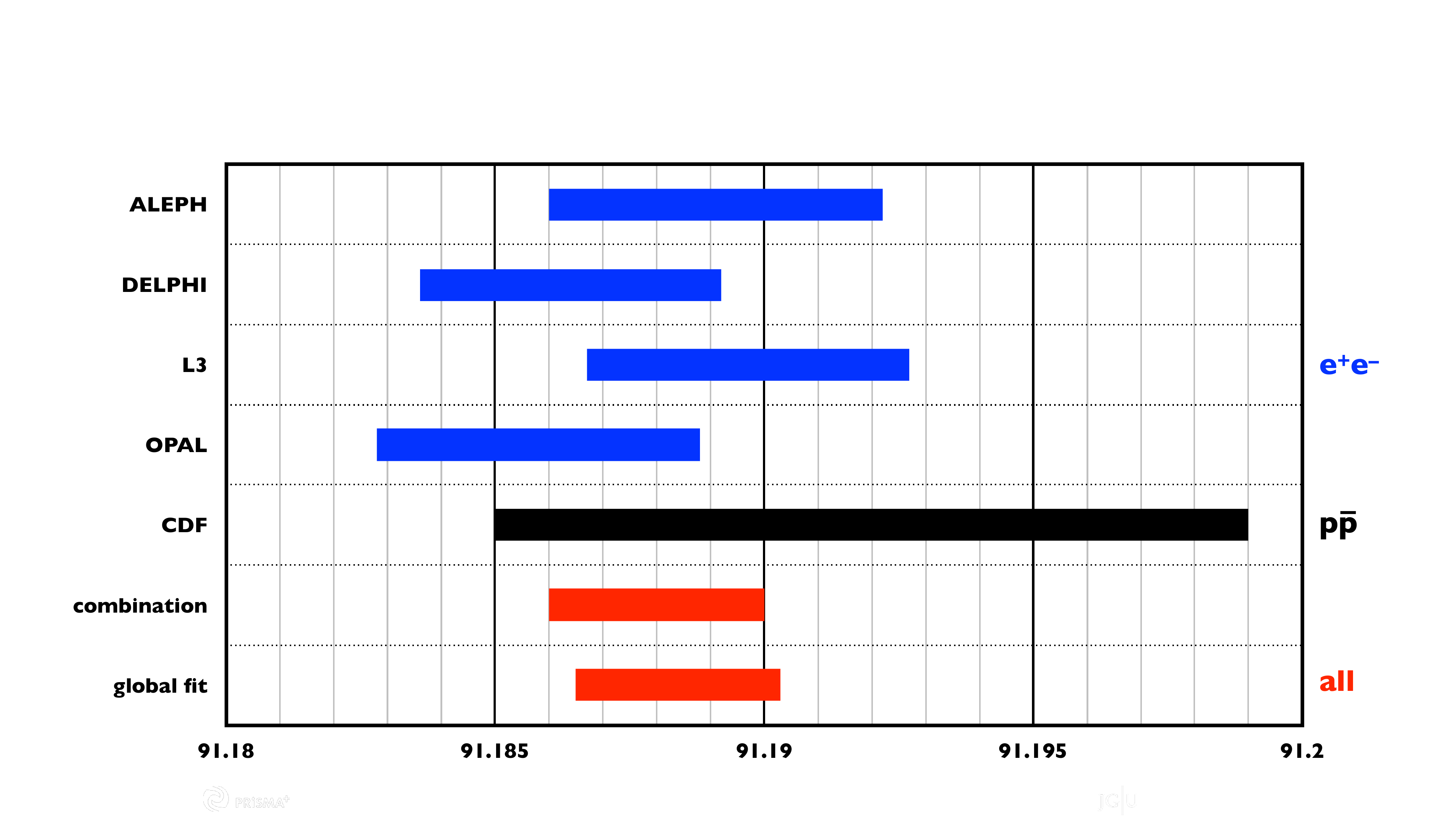}}
\end{minipage}
\caption[]{Determinations of $M_Z$ [GeV].
Shown are the measurements by the four individual experimental groups (general purpose detectors) at LEP (in blue),
and the result obtained by CDF (in black).
The combination and the result of the global EW fit are also shown (in red).}
\label{mz}
\end{figure}

\begin{figure}[t]
\begin{minipage}{\linewidth}
\centerline{\includegraphics[trim={0 0 0 180},clip,width=1.11\linewidth]{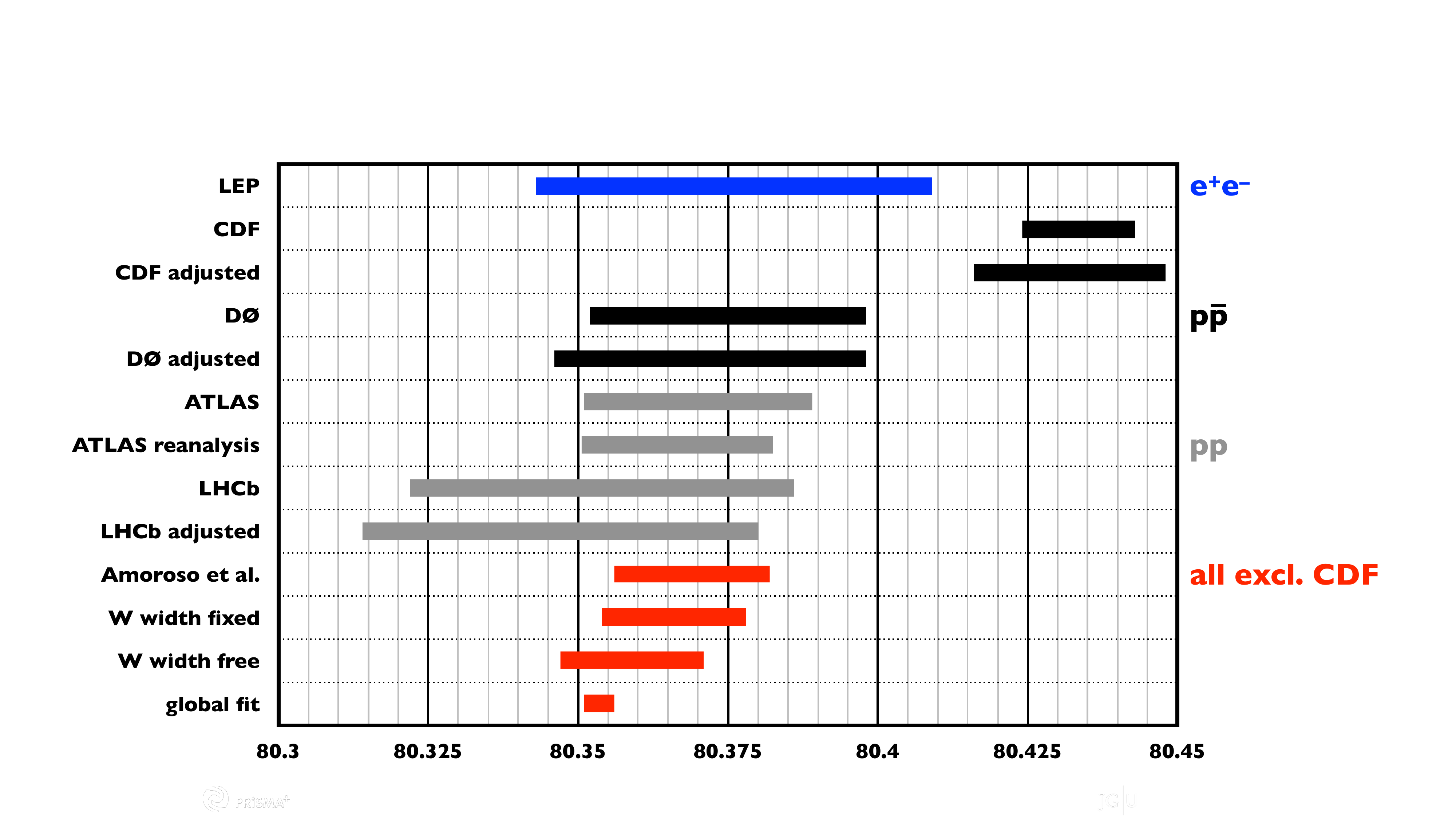}}
\end{minipage}
\caption[]{Determinations of $M_W$ [GeV]. 
The LEP~2 result~\cite{ALEPH:2013dgf} is the combination of the $W^*W^-$ threshold scan with the dominant kinematic reconstruction measurements 
(in blue).
The entries labeled CDF~\cite{CDF:2022hxs}, D\O~\cite{D0:2012kms}, ATLAS~\cite{ATLAS:2017rzl}, and LHCb~\cite{LHCb:2021bjt}, 
show the original results (see Ref.~\cite{CDF:2013dpa} for the Tevatron combination before the CDF reanalysis~\cite{CDF:2022hxs}), 
while the corresponding adjusted results~\cite{LHC-TeVMWWorkingGroup:2023zkn} are from the application 
of a coherent framework addressing theory uncertainties and correlations, including a common set of parton distribution functions.
The result of the very recent reanalysis~\cite{ATLAS:2024erm} of the ATLAS data from Run 1 at the LHC is also shown
which has been obtained by fixing the $W$ width, $\Gamma_W$.
If alternatively $\Gamma_W$ is allowed as a free parameter, their $M_W$ central value drops by 11.7~MeV.
The first of the combinations (in red) is from Ref.~\cite{LHC-TeVMWWorkingGroup:2023zkn}.
The second (third) combination assumes fixed (floating) values for $\Gamma_W$, while the last entry is from the global EW fit.}
\label{mw}
\end{figure}

These low $Q^2$-measurements of $\sin^2\theta_W$ in PVES~\cite{Kumar:2013yoa,Erler:2014fqa} are not only competitive with, 
but also highly complementary to the $Z$-pole determinations. 
This is illustrated in Figure~\ref{runningsin2tw} which shows the scale dependence of the weak mixing angle 
in the $\rm \overline{MS}$-renormalization scheme.
Around the $Z$ resonance, contributions from effective four-fermion amplitudes due to BSM physics are strongly suppressed,
but far below it they enter merely power-suppressed providing sensitivity to various types of new physics~\cite{Erler:2003yk},
such as a heavy $Z'$ boson~\cite{Erler:2009jh}.
Thus, the comparison of on- {\em vs.}~off $Z$-pole determinations could uncover such four-fermion amplitudes, 
which could mimic a departure of the SM running.
There may also be a parity-violating variant of a light dark photon --- called a dark $Z$ boson~\cite{Davoudiasl:2012qa} --- that could mix with the SM $Z$,
and reveal its presence by an actual change in the scale dependence setting in between the scales of different low-energy determinations.

The high $Q^2$-measurements of $\sin^2\theta_W$, on the other hand, are best suited to study possible modifications of the flavor-dependent
$Z$ couplings.
Finally, the whole suite of $\sin^2\theta_W$ measurements combined with $M_W, M_Z, \Gamma_Z$, 
and $G_F$ (determined from the $\mu$-lifetime~\cite{MuLan:2010shf}) can be used for the extraction of the so-called oblique parameters 
--- such as $S$, $T$, and $U$~\cite{Peskin:1990zt} --- constraining new physics contributions to gauge boson self-energies.

\section{Latest developments}
Recently, the CDF Collaboration~\cite{CDF:2022hxs} presented the first precision measurement of $M_Z$ at a hadron collider
from fits to the di-muon and di-electron invariant mass distributions. 
Our combination~\cite{ParticleDataGroup:2024cfk} of these two channels, $M_Z = 91.192 \pm 0.007$~GeV, 
combined with the LEP measurements~\cite{ALEPH:2005ab}, yields the new world average (see Figure \ref{mz}),
$$ M_Z = 91.1880 \pm 0.0020 \mbox{ GeV} $$

\begin{figure}[t]
\begin{minipage}{\linewidth}
\centerline{\includegraphics[trim={0 0 0 180},clip,width=1.11\linewidth]{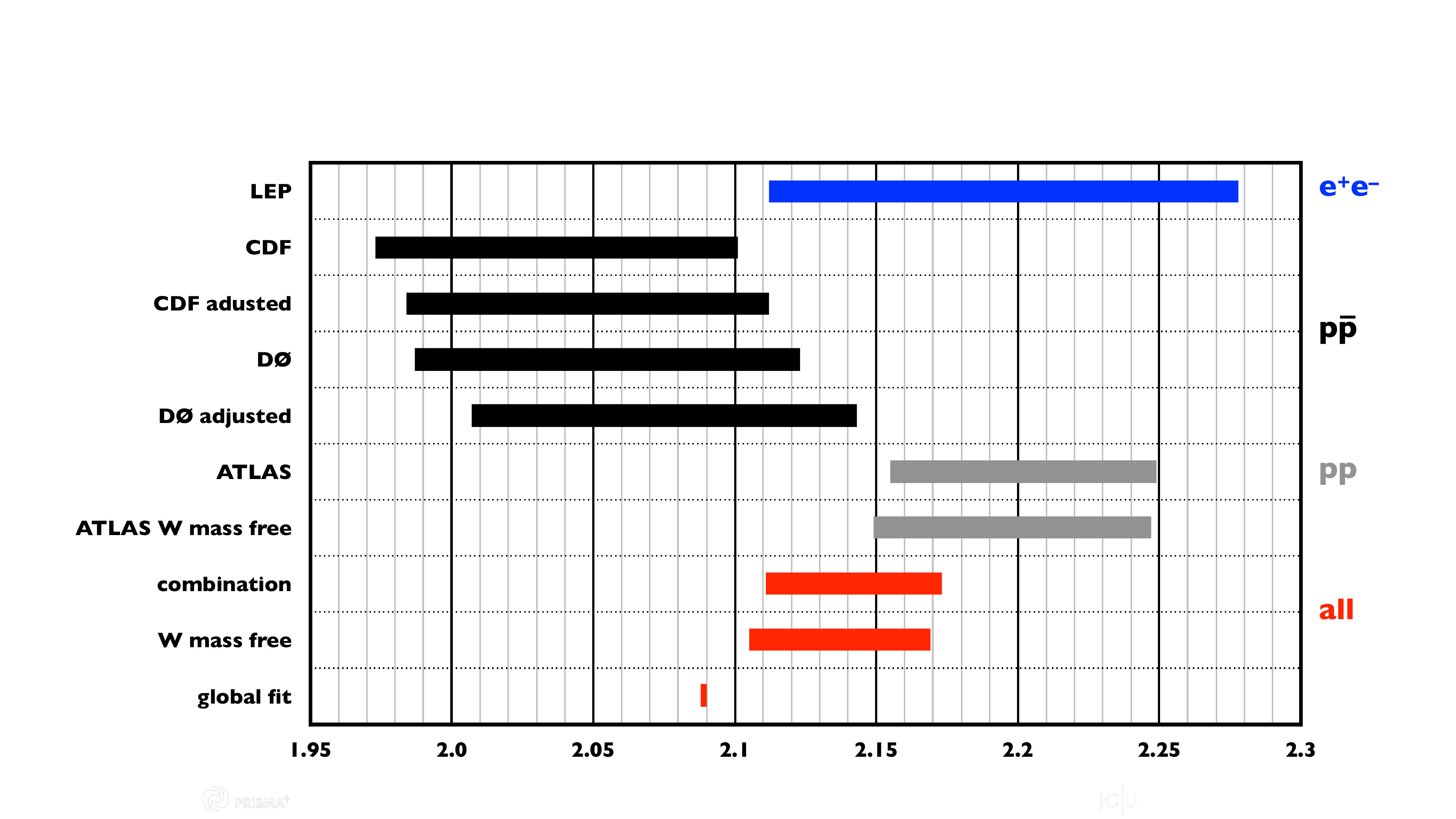}}
\end{minipage}
\caption[]{Determinations of $\Gamma_W$ [GeV].
The entries~\cite{TevatronElectroweakWorkingGroup:2010mao} labeled CDF and D\O\ show the original results.
Since these have been obtained relative to fixed and outdated values of $M_W$, 
we have also computed adjusted values using $M_W$ from ATLAS~\cite{ATLAS:2024erm} instead.
The first (second) of the combinations (in red) assumes fixed (floating) values for $M_W$, while the last entry is again from the global EW fit.}
\label{Wwidth}
\end{figure}

Figure~\ref{mw} shows the corresponding situation for $M_W$. 
As can be seen, the very precise determination by CDF~\cite{CDF:2022hxs} is incompatible with other measurements
which are in good agreement with each other and also with the global fit~\cite{ParticleDataGroup:2024cfk},
\be
   M_W = 80.356 \pm 0.005 \mbox{ GeV}
\ee
obtained when excluding the CDF result~\cite{CDF:2022hxs}.

Figure~\ref{Wwidth} summarizes the determinations of $\Gamma_W$.
The world averages of $M_W$ and $\Gamma_W$ excluding the CDF constraint 
(which is recommended by the LHC-TeV MW Working Group~\cite{LHC-TeVMWWorkingGroup:2023zkn}) are~\cite{ParticleDataGroup:2024cfk},
\begin{align}
           M_W &= 80.360 \pm 0.012 \mbox{ GeV} \\
\Gamma_W &= \ph0 2.136 \pm 0.032 \mbox{ GeV}
\end{align}
with a correlation coefficient of close to $-0.3$. 
Alternatively, fixing $\Gamma_W$ to the SM prediction gives 
\begin{equation}
   M_W = 80.366 \pm 0.012 \mbox{ GeV}
\end{equation} 
It is interesting that the determination at the forward detector LHCb is partially anti-correlated with those at ATLAS and CMS~\cite{Bozzi:2015zja},
which adds combinatorial power.

\begin{figure}[t]
\begin{minipage}{\linewidth}
\centerline{\includegraphics[trim={0 0 0 3},clip,width=1.04\linewidth]{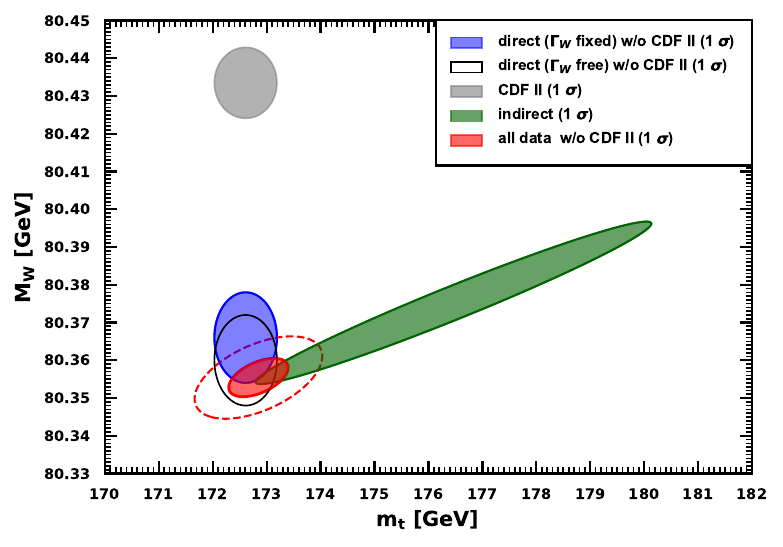}}
\end{minipage}
\caption[]{Current $1~\sigma$ constraints in $M_W$ {\em vs.}~$m_t$.
The purple ellipse is obtained by fixing $\Gamma_W$ to the SM prediction and excluding the CDF result~\cite{CDF:2022hxs}, while the open ellipse is for $\Gamma_W$ free.
The green contour corresponds to the indirect SM prediction by all other constraints.
The full (open dashed) red ellipse is the combination at $1~\sigma$ (90\% CL).
The CDF determination of $M_W$ (in gray) is in conflict.}
\label{mwmt}
\end{figure}

One can perform a global fit to all data excluding the direct measurements of the top quark mass at the hadron colliders.
The result,
\be
m_t = 175.2 \pm 1.8 \mbox{ GeV}
\ee
is $1.4~\sigma$ above the combination~\cite{ParticleDataGroup:2024cfk},
\be
m_t = 172.61 \pm 0.58 \mbox{ GeV}
\label{mt}
\ee
of many individual measurements at the Tevatron~\cite{CDF:2016vzt} and Run~1 at the LHC~\cite{ATLAS:2024dxp},
as well as a number of more recent results at Run~2~\cite{ParticleDataGroup:2024cfk}.
The error in Eq.~\eqref{mt} is dominated by the theoretical uncertainty from the conversion~\cite{Marquard:2015qpa} of the nominal results 
(which are assumed to be numerically close to the top quark pole mass)
to the short-distance definition in the $\rm \overline{MS}$ renormalization scheme, which actually enters the SM predictions and fits.
Figure~\ref{mwmt} displays the current $M_W$--$m_t$ parameter space.

\begin{figure}[t]
\begin{minipage}{\linewidth}
\centerline{\includegraphics[trim={0 0 0 180},clip,width=1.11\linewidth]{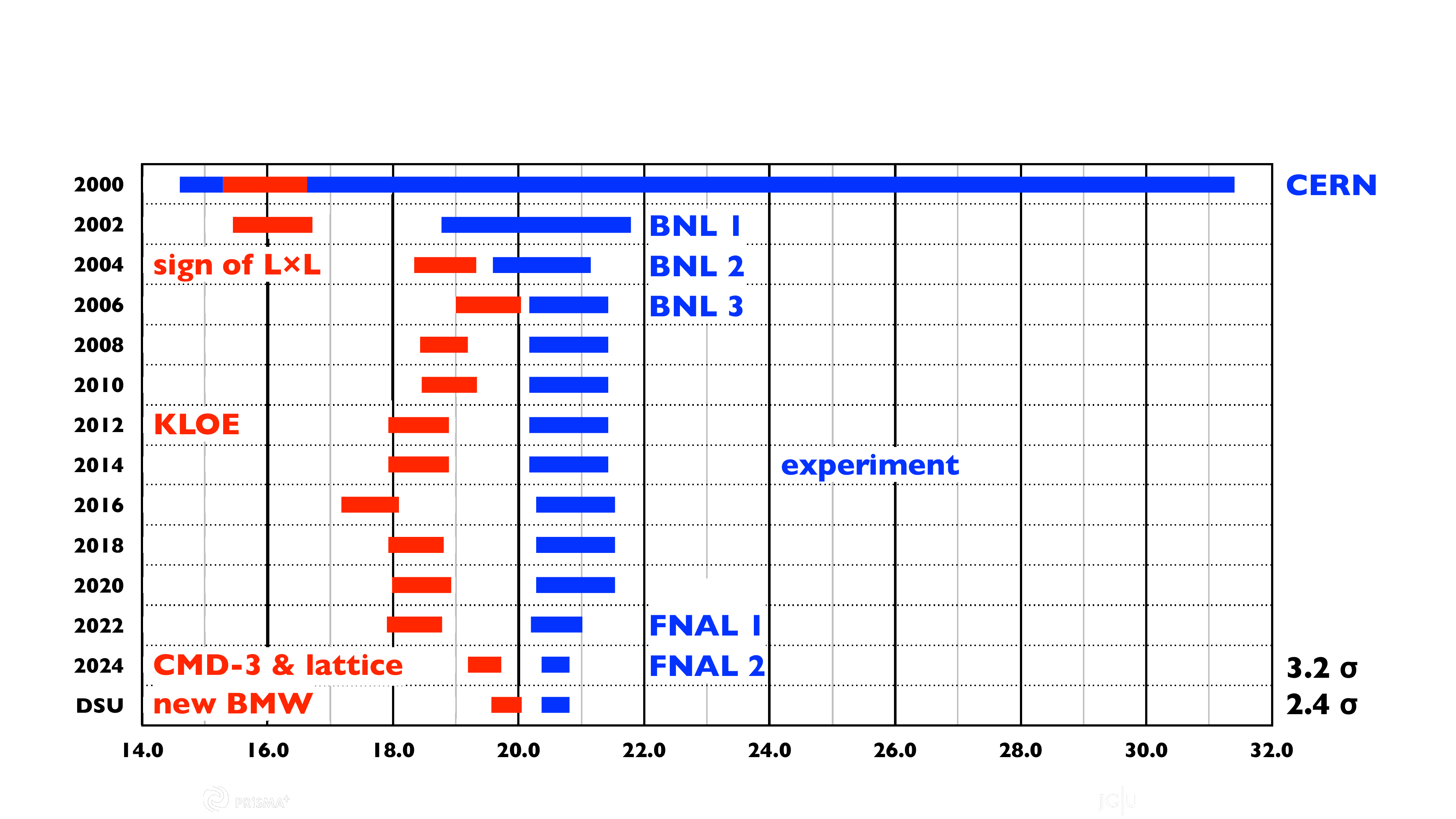}}
\end{minipage}
\caption[]{Predictions (in red) and measurements (in blue) of the muon anomalous magnetic moment $a_\mu$.
As can be seen, the experimental results from CERN~\cite{CERN-Mainz-Daresbury:1978ccd}, BNL~\cite{Muong-2:2006rrc}, 
and FNAL~\cite{Muong-2:2023cdq} have been remarkably stable over time, while the SM predictions varied well outside their nominal uncertainties.
The values indicated on the horizontal axis are for $10^9 a_\mu - 1165900$.}
\label{amu}
\end{figure}

Likewise, one can perform a fit to all data except for the direct determination of $M_H$ at the LHC, with the result,
\be
M_H = 97^{+18}_{-16} \mbox{ GeV}
\ee
which is $1.6~\sigma$ below the LHC combination~\cite{ParticleDataGroup:2024cfk},
\be
M_H = 125.10 \pm 0.09 \mbox{ GeV}
\label{MH}
\ee
of ATLAS~\cite{ATLAS:2023oaq} and CMS~\cite{CMS:2023ogh}, 
where we treated the smaller systematic error (from ATLAS) as common among the two determinations.

A fit allowing the aforementioned parameters $S$ and $T$~\cite{Peskin:1990zt} results in~\cite{ParticleDataGroup:2024cfk},
\begin{align}
           S &= -0.05 \pm 0.07 \\
           T &= \ph- 0.00 \pm 0.06
\end{align}
with a very strong correlation of 93\%.
A fit allowing only $T$ constrains new physics contributions to $\Delta\hat\rho$ in Eq.~\eqref{EWheart}.
For example, the mass splittings of hypothetical extra fermions are constrained by
\begin{equation}
(2 \hbox{ GeV})^2 < \sum_i \frac{N_C^i}{3}\, \Delta m^2_i < (44 \hbox{ GeV})^2
\label{Tpar}
\end{equation}
at the 90\% CL, where the sum runs over all new-physics doublets, such as fourth-family quarks and leptons or
vector-like fermion doublets which contribute to the sum in Eq.~\eqref{Tpar} with an extra factor of 2.

\begin{figure}[t]
\vspace{5pt}\hspace{0pt}
\begin{minipage}{\linewidth}
\centerline{\includegraphics[trim={0 0 8 0},clip,width=0.53\linewidth]{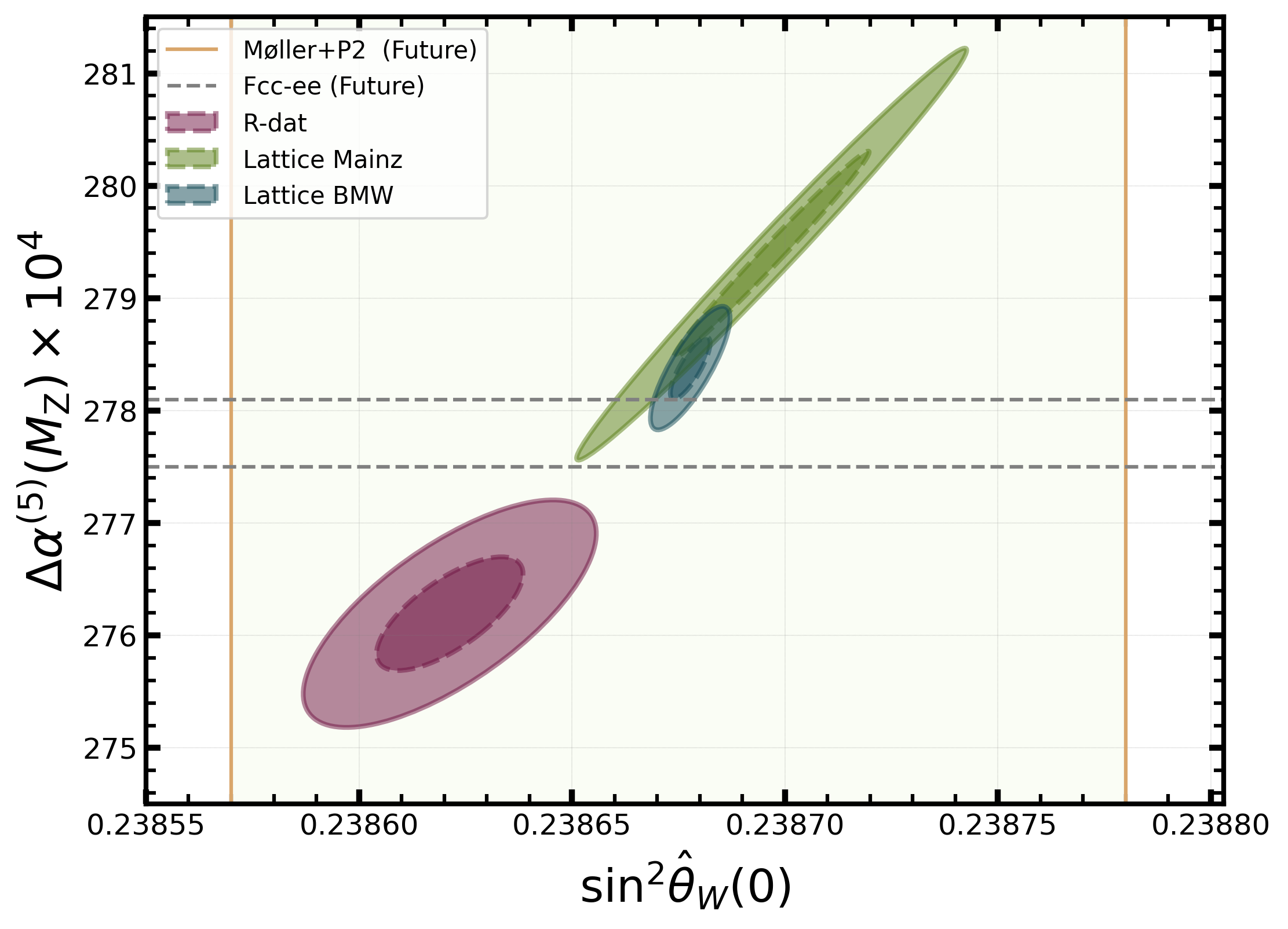}\hspace{-9pt}
\includegraphics[trim={30 0 0 0},clip,width=0.495\linewidth]{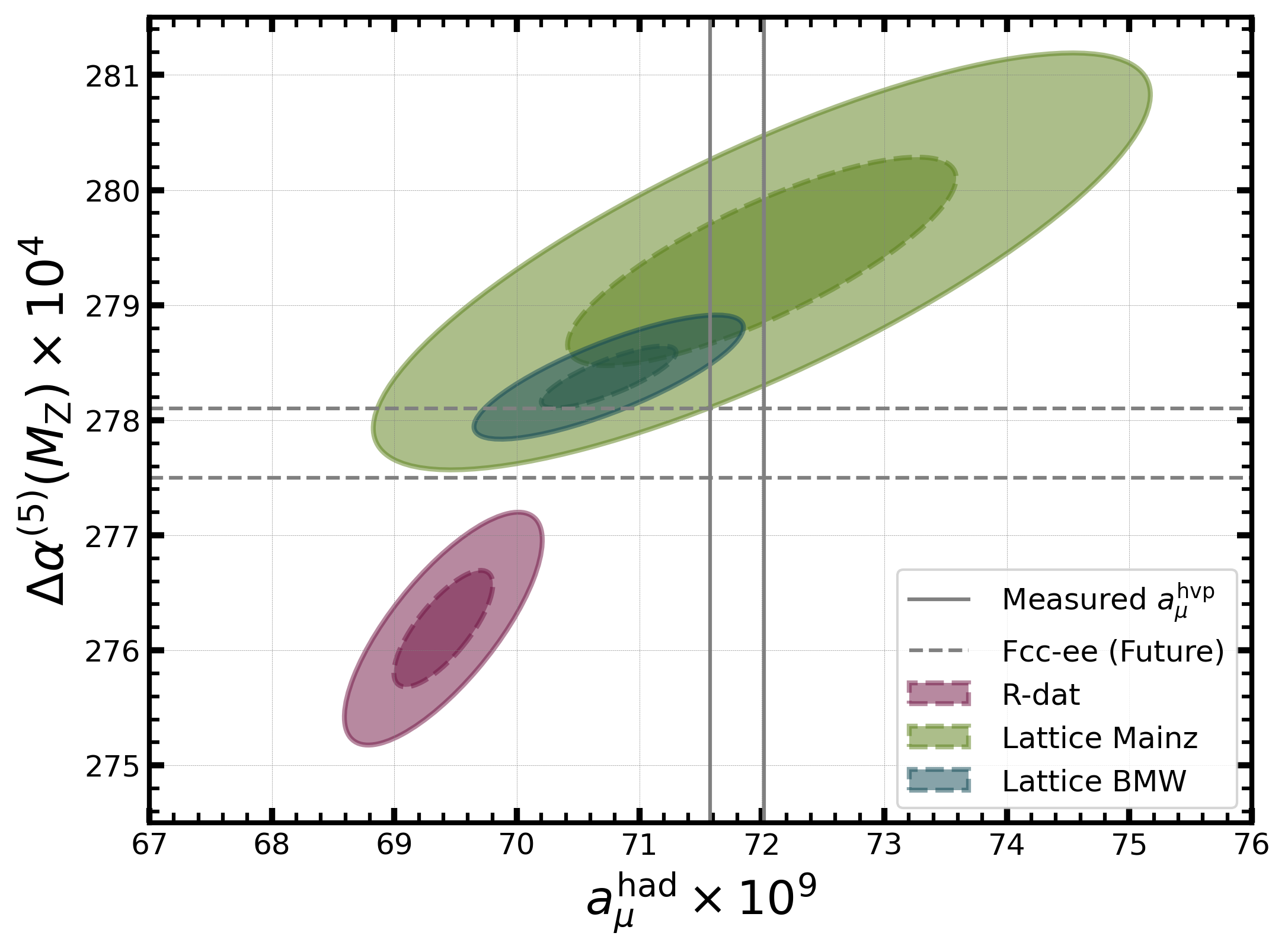}}
\vspace{-18pt}\hspace{193pt}\includegraphics[trim={0 0 0 0},clip,width=0.556\linewidth]{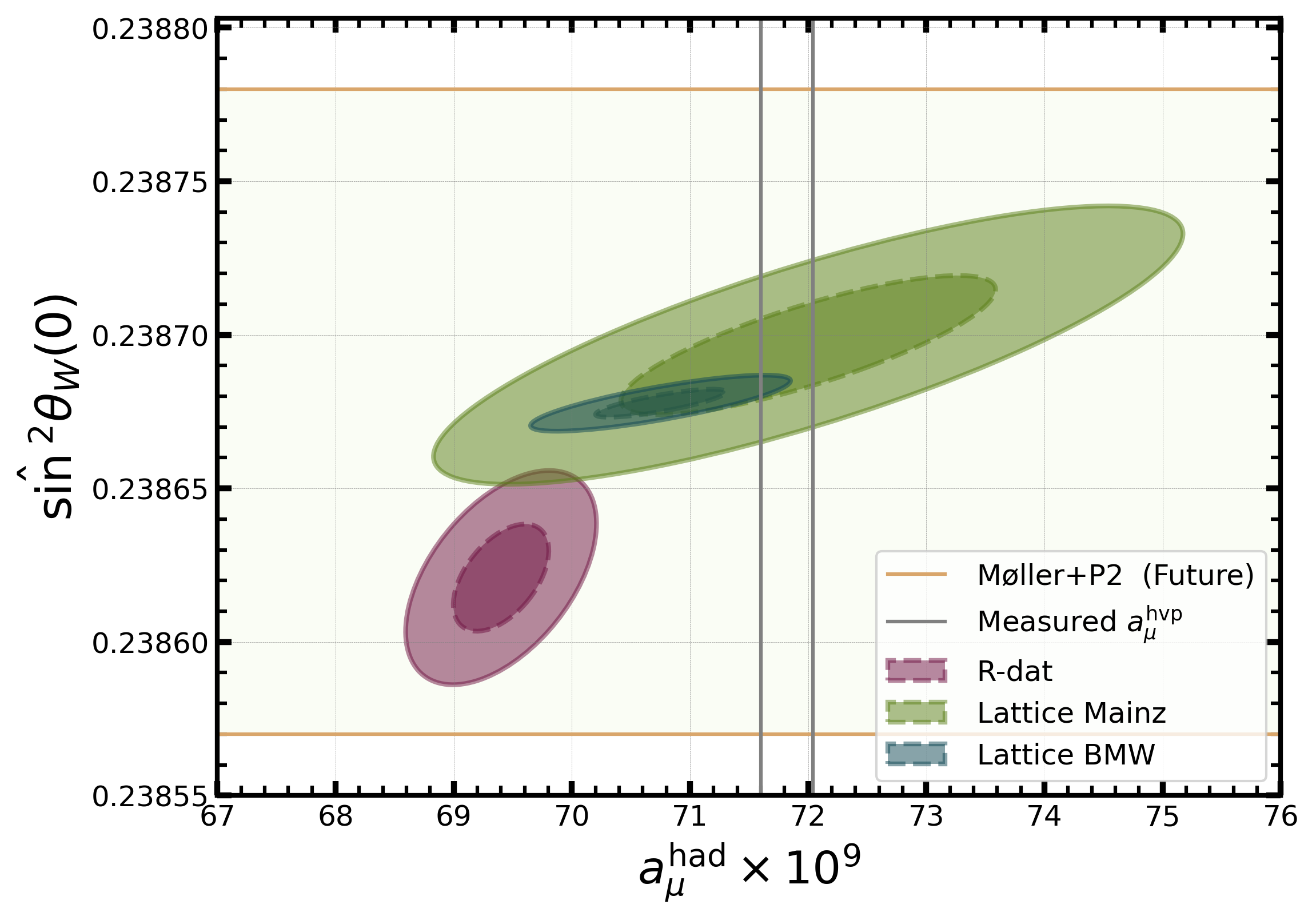}
\end{minipage}
\caption[]{Left: contours for $\Delta \chi^2 = 1$ (dashed) and $\Delta \chi^2 = 4$ (solid) in the $\Delta\alpha^{(5)}(M_Z)$ {\em vs.}~$\sin^2\hat\theta_W(0)$ 
plane, using $R$-ratio data (cherry red, lower left) and lattice results from BMW (blue, center) and Mainz~\cite{Ce:2022eix} (green, upper right) as inputs. 
The yellow vertical band represents the expected combined $1~\sigma$ range for P2 and MOLLER. 
The horizontal gray dashed band indicates a projection for the FCC-ee~\cite{FCC:2018evy}.
Upper right: the corresponding contours for $\Delta\alpha^{(5)}(M_Z)$ {\em vs.}~$a_\mu^{\rm had}$.
The width of the vertical band corresponds to the current experimental uncertainty in $a_\mu$.
Lower right: the corresponding contours for $\sin^2\hat\theta_W(0)$ {\em vs.}~$a_\mu^{\rm had}$.
(Plots reprinted from Ref.~\cite{Erler:2024lds}.)}
\label{confidence}
\end{figure}

\section{Hadronic vacuum polarization}
\label{HVP}
The largest radiative corrections to the EW precision observables are generally due to vacuum polarization effects,
where the hadronic part introduces a significant uncertainty. 
One of the most important such observable is the anomalous magnetic moment of the muon, $a_\mu$,
which has been measured to 0.19~ppm~\cite{Muong-2:2023cdq}.
The SM prediction, which is about 3~$\sigma$ below the experimental result,
is nominally of similar precision, where the uncertainty is dominated by the hadronic vacuum polarization contribution.
However, there is currently a large spread regarding the pertinent theoretical and experimental information,
where the latter enters into a dispersion integral over $Q^2$ of the cross section of $e^+ e^- \to $~hadrons and related quantities.
Indeed, the discrepancy is largely driven by the data collected with the KLOE detector at the Frascati $\phi$-factory DA$\Phi$NE~\cite{KLOE:2010qei},
while the recent result by the CMD-3 Collaboration at the VEPP-2000 $e^+ e^-$ collider at BINP in Novosibirsk~\cite{CMD-3:2023alj}
does not hint at any significant deviation. 
In turn, the data from BaBar~\cite{BaBar:2009wpw} and from hadronic $\tau$ decays~\cite{Davier:2023fpl} are somewhere in between these two.
The first competitive lattice gauge theory calculation~\cite{Borsanyi:2020mff} by the Budapest-Marseille-Wuppertal (BMW) Collaboration 
is in good agreement with $\tau$ decays, 
while their update~\cite{Boccaletti:2024guq} indicates good agreement between the SM prediction and the measurement of $a_\mu$.

The chronology of predictions and measurements of $a_\mu$ is shown in Figure~\ref{amu}.
The discrepancy that was seen back in 2002 was due to a sign error in the evaluation of the hadronic light-by-light contribution, $a_\mu^{\rm LBL}$,
which was indeed opposite to what one should have expected from the corresponding leptonic effects\footnote{The results 
from the leptonic sector can also be used to argue for an upper limit, $a_\mu^{\rm LBL} \lesssim 1.59 \times 10^{-9}$ (95\% CL)~\cite{Erler:2006vu}.}.
The significant shift in 2024 was due to both, the results by BMW~\cite{Borsanyi:2020mff} and CMD-3~\cite{CMD-3:2023alj}.
The further shift --- evaluated for this conference --- is due to the very recent BMW update~\cite{Boccaletti:2024guq} quoting a 40\% smaller uncertainty.

The hadronic vacuum polarization effects enter other observables, as well, introducing theoretical correlations among otherwise unrelated measurements.
In particular, the quark contribution (excluding the top) $\Delta\alpha^{(5)}(M_Z)$ to the running of $\alpha$ to the scale $M_Z$ 
enters $\Delta\hat r$ in Eq.~\eqref{EWheart}, and is needed, {\em e.g.\/} for the calculation of $M_W$ and the extraction of $M_H$ from loop effects.
Likewise, the hadronic contribution to the running of $\sin^2\hat\theta_W$ is crucial for the interpretation of experiments 
dedicated to measure $A_{LR}$ in Eq.~\eqref{ALR}.
The induced correlations between these quantities are illustrated in Figure~\ref{confidence}.
They have been obtained --- for the first time --- purely theoretically, \ie by combining QCD perturbation theory with lattice QCD 
in the non-perturbative regime~\cite{Ce:2022eix}.

It is interesting that replacing the data-driven result in $\Delta\alpha^{(5)}(M_Z)$ of 2022 with the more theory-driven one in 2024
lowers the SM prediction of $M_W$ by $2.7$~MeV and the loop prediction of $M_H$ by $7.0$~GeV.

\section{Conclusions and outlook}
After more than 50 years of electroweak precision physics, there is still no conclusive evidence for BSM physics.
Meanwhile, EW precision tests continue to be of crucial importance, and it is worth reminding that the masses
$m_c$, $M_W$, $M_Z$, $m_t$, and $M_H$ have all been successfully predicted {\em before\/} their discoveries.
The quality of the global fit is very good with a $\chi^2$ per degree of freedom of $49.5/47$, corresponding to a 37\% probability for a larger $\chi^2$.
Averaging in the latest result by the BMW Collaboration~\cite{Boccaletti:2024guq} significantly reduced the discrepancy between
the theoretical prediction and measurement of $a_\mu$ which is now\footnote{After the conclusion of the DSU conference a new
lattice calculation~\cite{Djukanovic:2024cmq} of $a_\mu^{\rm had}$ indicates even better agreement with the experimental result.} 
at a moderate $2.4~\sigma$.
The CDF Collaboration reported on very precise measurements of $M_Z$ and $M_W$,
where $M_Z$ is in good agreement with LEP, while $M_W$ is nominally about 7~$\sigma$ higher than the other measurements.

In the near future, we will see the advent of ultra-high EW precision measurements at low~$Q^2$,
which will represent competitive alternatives to the high energy frontier.
And of course, a spectacular leap in precision can be expected from a future lepton collider~\cite{Fan:2014vta}, such as the ILC~\cite{ILC:2007bjz}, 
the CEPC~\cite{CEPCStudyGroup:2018ghi}; the FCC-ee~\cite{FCC:2018evy}, or a muon collider~\cite{Accettura:2023ked}.

\acknowledgments{Many thanks to the organizers of DSU 2024 for an unforgettable conference and for introducing us to the culture and food of Corfu.
I would also like to sincerely thank my recent collaborators on these topics, especially Rodolfo Ferro, Ayres Freitas and Simon Kuberski.}

\end{document}